\documentclass[12pt]{iopart}
\usepackage[]{graphicx}
\newcommand*{\be}{\begin{equation}}
\newcommand*{\ee}{\end{equation}}
\begin{document}
\title[Modulational instability and nonlocality management
in coupled NLS system]{Modulational instability and nonlocality management in coupled NLS system}
\author{Evgeny V Doktorov and Maxim A Molchan}
\address{ B I Stepanov Institute of Physics, 220072 Minsk, Belarus}
\eads{\mailto{doktorov@dragon.bas-net.by},
\mailto{m.moltschan@dragon.bas-net.by}}

\begin{abstract}
The modulational instability of two interacting waves in a
nonlocal Kerr-type medium is considered analytically and
numerically. For a generic choice of wave amplitudes, we give a
complete description of stable/unstable regimes for zero
group-velocity mismatch. It is shown that nonlocality suppresses
considerably the growth rate and bandwidth of instability. For
nonzero group-velocity mismatch we perform a geometrical analysis
of a nonlocality management which can provide stability of waves
otherwise unstable in a local medium.
\end{abstract}

 \submitto{Physica Scripta}
\maketitle

\section{Introduction}
Various nonlinear dispersive wave systems exhibit instability,
known as the modulational instability (MI). MI stems from the
interaction between nonlinearity and group-velocity dispersion and
manifests itself as a self-induced amplitude modulation of a
continuous wave propagating in a nonlinear medium, with subsequent
generation of localized structures. MI arises in many physical
settings including plasma~\cite{rev1,rev2,rev3,rev4},
fluids~\cite{rev5}, nonlinear optics~\cite{rev6,rev7,rev8,rev9},
and atomic Bose-Einstein condensates~\cite{rev10,rev11}.

In the context of nonlinear fibre optics, MI requires anomalous
dispersion of a medium to exist. In this case, negative
group-velocity dispersion combined with the self-phase modulation
amplifies the modulational frequency sidebands producing a train
of ultra-short pulses as a result of breakup of a continuous
wave~\cite{rev12,rev8,rev12a}.

On the other hand, when two or more optical beams propagate down
the fibre, MI can occur even in the normal-dispersion regime, at
the cost of the cross-phase modulation~\cite{rev13,rev14}. Such a
situation is usually described by a set of coupled nonlinear
Schr\"odinger (NLS) equations. A thorough analysis of MI for two
nonlinearly interacting waves in a Kerr medium was performed in
recent papers~\cite{rev15,rev16}.

The above results concern the class of local nonlinear media when
the response of the medium at a particular point depends solely on
the wave intensity at that point. Alternatively, nonlocality plays
a key role in physical systems where transport phenomena and
finite-range interaction cannot be neglected. The examples of
importance of spatial nonlocality for the development of MI can be
found in plasmas~\cite{rev17,rev18}, liquid crystals~\cite{rev19},
discrete nonlinear systems~\cite{rev20}, Bose-Einstein
condensates~\cite{rev21}. In particular, nonlocality changes
drastically the growth rate and bandwidth of instability caused by
stochasticity of parameters of a Kerr medium~\cite{rev22}. The
present status of MI of scalar waves in nonlocal media is
discussed in the review papers~\cite{rev23,rev24}.

In our paper we extend the analysis of MI in coupled wave systems
to the case of nonlocal media. It was demonstrated recently that
vector soliton structures in nonlocal media exhibit properties
that have no counterpart in the scalar case~\cite{rev25}. In
particular, a possibility was revealed to stabilize multipole
(dipole, quadrupole, hexapole, etc.) vector solitons in bulk
nonlocal media at the cost of vectorial coupling \cite{rev27a}. In
this case mutually incoherent nodeless and multipole components
jointly induce a nonlinear refractive index profile. As regards
higher-order nonlinearities, a one-dimensional phenomenological
model of a nonlocal medium with focusing cubic and defocusing
quintic nonlinearities was developed in \cite{rev28a}. Two types
of solitons, even-parity (fundamental) and odd-parity (dipole)
were found and their stability was explored. Note that solitons of
this sort are unstable in local media. Lastly, an interesting
result should be mentioned concerning formation of stable
three-dimensional spatiotemporal solitons in nonlocal Kerr media
\cite{rev29a}. Fundamental soliton of this model is stable
provided its propagation constant exceeds a certain critical value
inverse proportional to a nonlocality parameter.

Below we derive the dispersion relation which characterizes the
propagation of two interacting waves in a nonlocal Kerr-type
medium. For the case of zero group-velocity mismatch we are able
to perform a complete analytic description of
stability/instability regimes for generic wave amplitudes. It is
shown that nonlocality suppresses considerably the growth rate and
bandwidth of instability. For nonzero velocity mismatch, we
investigate a role of nonlocality in a management of the stability
properties in the system under consideration. Nonlocality is shown
to provide an efficient stabilization mechanism permitting
stabilize interacting waves which are strongly unstable in a local
medium.

\section{Model}

Propagation of two nonlinearly interacting waves with complex
amplitudes $u(t,x)$ and $v(t,x)$ in a nonlocal Kerr medium is
modelled by the following system of dimensionless equations:
 \begin{eqnarray}\label{CEq}
\eqalign{ \fl \rmi\left(\frac{\partial u}{\partial
t}+V_1\frac{\partial u}{\partial x}\right)
+ \frac{d_1}{2}\frac{\partial^2u}{\partial x^2}\\
+s_1u \int\rmd x'R(x-x')\left(|u(t,x')|^2+\alpha|v(t,x')|^2\right)=0, \\
 \fl \rmi\left(\frac{\partial v}{\partial t}+V_2\frac{\partial v}{\partial x}\right)
 + \frac{d_2}{2}\frac{\partial^2v}{\partial x^2}\\
+s_2v \int\rmd
x'R(x-x')\left(|v(t,x')|^2+\alpha|u(t,x')|^2\right)=0,}
\end{eqnarray}
which naturally generalize those \cite{rev25a} for a local medium.

Physical meaning of coordinates $t$ and $x$ depends on the context
of the problem: they are time and coordinate for Bose-Einstein
condensates, longitudinal coordinate $z$ and transversal
coordinate $x$ for optical beams in waveguides, the propagation
coordinate $z$ and time $t$ for optical pulses in fibres,
respectively. The parameters $d_j$ ($j=1,2$) determine effective
atom masses in condensate models, diffraction coefficients in
waveguides and group-velocity dispersion for pulses. The
parameters $V_j$ stand for group velocities, $s_j$ determine
nonlinearity coefficients, and $\alpha>0$ is a cross-phase
modulation parameter. No restrictions are posed on signs and
magnitudes of $d_j$ and $s_j$. $R(x)$ is a positively definite
symmetric response function of a nonlinear medium obeying the
normalization condition $\int\rmd xR(x)=1$ (we do not specify the
integration limits when the integration is carried out along the
whole line). An important development of nonlocal models with
asymmetric response functions relevant, e.g., to account for the
Raman effect, was achieved by Wyller~\cite{rev26}.

We will illustrate our results by the exponential response function
\be \label{exp}
R(x)=\frac{1}{2\sigma}\exp\left( -\frac{|x|}{\sigma}\right).
\ee
Here $\sigma$ is a nonlocality parameter.
The function (\ref{exp}) occurs as a solution of the diffusion-type equation for
nonlinearity~\cite{rev25} and adequately describes the nonlinear response of
thermo-optical materials, liquid crystals and partially ionized plasmas. For
$\sigma\to 0$ we have $R(x)\to \delta (x)$ and hence reproduce the local model
used in~\cite{rev16}.

\section{Modulational instability: general analysis}

In order to investigate the modulational instability in the system
(\ref{CEq}), we start with the continuous wave solution of the
form \be \label{solutions} u=u_0\rme^{-\rmi\omega_1 t},\qquad
v=v_0\rme^{-\rmi\omega_2 t}, \ee where $u_0$ and $v_0$ are
constant real amplitudes, and frequencies $\omega_j$ are given by
the relations
\[
\omega_1=-s_1\left(u_0^2+\alpha v_0^2\right),\qquad \omega_2=-s_2\left(\alpha
u_0^2+v_0^2\right),
\]
which are the same as for the local coupled NLS system.

Following the standard procedure, let us consider a small
perturbation of the continuous waves (\ref{solutions}):
\be
\label{eq4} u=(u_0+a(t,x))\rme^{-\rmi\omega_1 t},\qquad
v=(v_0+b(t,x))\rme^{-\rmi\omega_2 t}.
\ee
Substituting (\ref{eq4})
into (\ref{CEq}) and neglecting terms with the second and higher
orders of the complex perturbation amplitudes $a$ and $b$, we
obtain a system of linear equations
 \begin{eqnarray}\label{eq5}
\eqalign{ \fl \rmi\left(\frac{\partial a}{\partial
t}+V_1\frac{\partial a}{\partial x}\right) +
\frac{d_1}{2}\frac{\partial^2a}{\partial x^2}+
s_1u_0^2 \int\rmd x'R(x-x')(a+a^*)(t,x')\\
+\alpha s_1u_0v_0 \int\rmd x'R(x-x')(b+b^*)(t,x')=0, \\
 \fl \rmi\left(\frac{\partial b}{\partial t}+V_2\frac{\partial b}{\partial x}\right)
 + \frac{d_2}{2}\frac{\partial^2b}{\partial x^2}+
s_2v_0^2 \int\rmd x'R(x-x')(b+b^*)(t,x')\\
+\alpha s_2u_0v_0 \int\rmd x'R(x-x')(a+a^*)(t,x')=0.}
\end{eqnarray}
We seek for solutions of (\ref{eq5}) in the form
\begin{eqnarray}
\nonumber a=A_+\rme^{\rmi(\kappa x-\Omega t)}+A_-^*\rme^{-\rmi(\kappa x-\Omega^* t)},\\
\nonumber b=B_+\rme^{\rmi(\kappa x-\Omega
t)}+B_-^*\rme^{-\rmi(\kappa x-\Omega^* t)},
\end{eqnarray}
where $\kappa$ and $\Omega$ are the wave number and frequency of
the perturbation, respectively. Hence, the condition of
solvability of the system (\ref{eq5}) takes the form of the
dispersion relation
\be\label{eq6}
\left[\left(\Omega-V_1\kappa\right)^2-\Omega_1^2 \right]
\left[\left(\Omega-V_2\kappa\right)^2-\Omega_2^2
\right]-\Omega_{12}^4=0,
\ee
where
\be\label{eq7}
\eqalign{\fl
\Omega_1^2=\frac{1}{4}d_1\kappa^2\left(d_1\kappa^2-4s_1u_0^2\hat
R\right),\qquad
\Omega_2^2=\frac{1}{4}d_2\kappa^2\left(d_2\kappa^2-4s_2v_0^2\hat R\right),\\
\Omega_{12}^4=d_1d_2s_1s_2\left(\alpha u_0v_0\kappa^2\hat
R\right)^2.}
\ee
Note the appearance of the Fourier transform
$\hat R(\kappa)$ of the response function $R(x)$,
\be \label{eq8}
\hat R(\kappa)=\int\rmd x\, R(x)\rme^{\rmi\kappa
x}=\frac{1}{1+\sigma^2\kappa^2},\qquad \hat R(-\kappa)=\hat
R^*(\kappa),
\ee
in the expressions for $\Omega_j^2$ and
$\Omega_{12}^4$. Since $\hat R(\kappa)\to 1$ for $\sigma\to 0$,
the expression~(\ref{eq6}) reduces in this limit to the well-known
dispersion relation for the local model~\cite{rev14,rev15,rev16}.

The dispersion relation (\ref{eq6}) takes the form of a general fourth-order
algebraic equation for $\Omega$. In order to provide better insight into the
properties of its solutions, we begin the analysis from the particular case of
equal group velocities, $V_1=V_2$, which allows us to perform the complete analytical
treatment.

\section{Equal group velocities}

If $V_1=V_2\equiv V$, we can eliminate the terms with $V_j$ from
(\ref{eq5}) by the Galilean transformation $t'=t$, $x'=x-Vt$, and
the dispersion relation (\ref{eq6}) reduces to the biquadratic
polynomial in $\Omega$,
\be \label{eq9}
\left(\Omega^2-\Omega_1^2
\right)\left(\Omega^2-\Omega_2^2 \right)-\Omega_{12}^4=0.
\ee
Introducing the definitions
\begin{eqnarray}\label{eq10}
\eqalign{\mu_j=s_jd_j,\qquad \varphi^2=\frac{1}{4}(d_1^2+d_2^2),\\
\psi=\mu_1u_0^2+\mu_2v_0^2,\qquad N(\kappa)=\kappa^{-2}\hat
R(\kappa),}
\end{eqnarray}
we write equation (\ref{eq9}) in the form
\be\label{eq11}
\eqalign{\fl\Omega^4-\kappa^4\left(\varphi^2-N\psi\right)\Omega^2+\\
\frac{\kappa^2}{16}\left[d_1^2d_2^2-4N\left(\mu_1d_2^2u_0^2+\mu_2d_1^2v_0^2\right)
+16N^2(1-\alpha^2)\mu_1\mu_2u_0^2v_0^2\right]=0.} \ee In the local
models, stability properties are determined directly by the
analysis of the function $\Omega(\kappa)$. In contrast, to
establish general properties of the stability/instability domains
for the nonlocal case, we will primarily consider the function
$\Omega(N)$. More detailed information will be then provided by a
particular choice of the positively definite function $N(\kappa)$.

 Solving (\ref{eq11}) gives
 \be \label{eq12}
 \Omega_\pm^2=\frac{1}{2}\kappa^4\left(\varphi^2-N\psi\pm
 \sqrt{D}\right),
 \ee
 where \be \label{eq13}
 D(N)=4\Delta_1N^2+4\delta\eta N+\delta^2 \ee and \be \label{eq14}
 \eqalign{\fl \Delta_1=\mu_1\mu_2(\alpha u_0v_0)^2+\eta^2,\qquad
 \eta=\frac{1}{2}
 \left(\mu_1u_0^2-\mu_2v_0^2\right),\\
 \delta=\frac{1}{4}\left(d_1^2-d_2^2\right).}
 \ee
 Without loss of generality we take $\delta\ge0$.

Stability of continuous waves (\ref{solutions}) will be ensured if
both $\Omega_+^2$ and $\Omega_-^2$ are positive for any choice of
the continuous wave amplitudes $u_0$ and $v_0$. In other words, we
do not consider situations when stability/instability conditions
fulfil only for specific relations between $u_0$ and $v_0$,
because they are not generally satisfied. Evidently, we get
$\Omega_\pm^2>0$ if $D>0$ and $\sqrt{D}<\varphi^2-N\psi$.

\subsection{Analysis}
To reveal positiveness regions of $\Omega_\pm^2$, we analyze at
first the function $D(N)$ (\ref{eq13}). Positiveness of $D(N)$
depends on $\Delta_1$ and $\eta$. It is seen from (\ref{eq14})
that $\Delta_1>0$ for any $\alpha$, $u_0$ and $v_0$ if
$\mu_1\mu_2>0$, as well as for $\alpha<1$ if $\mu_1\mu_2<0$. Note
that $\Delta_1$ could be positive for $\mu_1\mu_2<0$ and
$\alpha>1$ under a specific choice of $u_0$ and $v_0$, but we
disregard such a possibility, as was stated above. Hence, in what
follows we consider $\Delta_1$ to be positive for generic choice
of $u_0$ and $v_0$, and for only restriction $\alpha<1$ if
$\mu_1\mu_2<0$.

As regards the role of $\eta$, it can be elucidated from the
analysis of the discriminant $D_1=-16\mu_1\mu_2(\alpha\delta
u_0v_0)^2$ of the equation $D(N)=0$. Evidently, $D>0$ if $D_1<0$,
i.e., for $\mu_1\mu_2>0$ and arbitrary $\alpha$ (the parabola
$D(N)=0$ lies above the horizontal axis $N$). If $D_1>0$ that
occurs for $\mu_1\mu_2<0$ and hence for $\alpha<1$ (two real
intersection points of the parabola with the $N$ axis), we have
two possibilities for $D>0$, depending on the sign of $\eta$:
\begin{enumerate}
\item $\eta>0$ means $\mu_1<0$ and $\mu_2<0$. Both intersection points lie on the negative
(unphysical) part of the $N$ axis, and $D>0$ for any $N>0$;

\item $\eta<0$ means $\mu_1<0$ and $\mu_2>0$. Hence, $D>0$ for $N\in(0, N'_-)
\bigcup(N'_+,\infty)$,
\end{enumerate}
where
\be \label{eq15}
N'_\pm=\frac{\delta}{2\Delta_1}\left(\pm\sqrt{\eta^2-\Delta_1}+|\eta|\right)>0.
\ee
Lastly, $D<0$ if $D_1>0$, $\alpha<1$, $\eta<0$, and $N\in (N'_-,N'_+)$. Let us
summarize the information obtained from the above analysis for the generic $u_0$
and $v_0$:

$D>0$ for $\Delta_1>0$ and
\begin{enumerate}
\item $\mu_1\mu_2>0$, $\forall (\alpha,N)$;

\item $\mu_1>0$, $\mu_2<0$, $\alpha<1$, $\forall N$;

\item $\mu_1<0$, $\mu_2>0$, $\alpha<1$, $N\in(0, N'_-)\bigcup(N'_+,\infty)$;\\

$D<0$ for $\Delta_1>0$ and

\item $\mu_1<0$, $\mu_2>0$, $\alpha<1$, $N\in (N'_-,N'_+)$.
\end{enumerate}

To study the condition $\sqrt{D}<\varphi^2-N\psi$, we introduce
$Z=D-(\varphi^2-N\psi)^2$, or
\be \label{eq16}
Z=4\Delta_2N^2+4fN+g,
\ee
where
\[
\Delta_2=(\alpha^2-1)\mu_1\mu_2u_0^2\mu_2^2,\qquad f=\frac{1}{4}\left(\mu_1d_1^2u_0^2
+\mu_2d_2^2v_0^2\right),\\
\]

\[
g=\frac{1}{4}d_1^2d_2^2.
\]
They are negative values of $Z$ that correspond to the condition
$\sqrt{D}<\varphi^2-N\psi$, and we require $\varphi^2>N\psi$.

The discriminant $D_2=16(f^2-\Delta_2g)$ of the equation $Z(N)=0$ has the form
\[
D_2=\left(\mu_1d_1^2u_0^2+\mu_2d_2^2v_0^2\right)^2-4(1-\alpha^2)\mu_1\mu_2
\left(d_1d_2u_0v_0\right)^2.
\]
It is seen that $D_2<0$ occurs only for special requirements on $u_0$ and $v_0$;
hence, we take in what follows $D_2>0$ for generic $u_0$ and $v_0$. This condition
takes place for $\mu_1\mu_2>0$ and any $\alpha$, as well as for $\mu_1\mu_2<0$
and $\alpha<1$.

Now we can establish properties of $Z(N)$. First of all we see that the presence
of $f$ in (\ref{eq16}) excludes the case $\mu_1\mu_2<0$ because it would lead to
nongeneric conditions on $u_0$ and $v_0$. Solutions of $Z=0$ are written as
\[
Z_\pm=\frac{1}{2\Delta_2}\left(\pm\sqrt{f^2+\frac{1}{4}d_1^2d_2^2\Delta_2}-f\right).
\]
Since $D_2>0$, both $Z_+$ and $Z_-$ are real. At first we consider
the case $\Delta_2>0$ which occurs for $\mu_1\mu_2>0$ and
$\alpha>1$. If $f>0$, i.e. $\mu_1>0$, $\mu_2>0$, and hence
$\alpha>1$, we obtain $Z_+>0$ and $Z_-<0$. Therefore, $Z$ will be
positive for $N>Z_+$ and negative for $N\in(0,Z_+)$. For $f<0$,
i.e., $\mu_1<0$, $\mu_2<0$ and $\alpha>1$, the solutions of
$Z(N)=0$,
\be \label{eq17}
Z'_\pm=\frac{1}{2\Delta_2}\left(\pm\sqrt{f^2+\frac{1}{4}d_1^2d_2^2\Delta_2}+|f|\right)
\ee
obey the same properties as $Z_\pm$, i.e., $Z>0$ for $N>Z'_+$
and $Z<0$ for $N\in(0,Z'_+)$.

If $\Delta_2<0$ (i.e., $\mu_1\mu_2>0$ and $\alpha<1$), both roots of $Z(N)=0$ are
positive for $f>0$ ($\mu_1>0$, $\mu_2>0$),
\[
\tilde{Z_\pm}=\frac{1}{4|\Delta_2|}\left(d_1^2u_0^2+d_2^2v_0^2\pm\sqrt{
\left(d_1^2u_0^2+d_2^2v_0^2\right)^2-
|\Delta_2|d_1^2d_2^2}\right)>0,
\]
and negative for $f<0$ ($\mu_1<0$ and $\mu_2<0$),
\[
\tilde{Z'_\pm}=-\frac{1}{4|\Delta_2|}\left(d_1^2u_0^2+d_2^2v_0^2\pm\sqrt{
\left(d_1^2u_0^2+d_2^2v_0^2\right)^2-
|\Delta_2|d_1^2d_2^2}\right)<0.
\]
Hence, for $\Delta_2<0$ we have $Z(N)>0$ for $f>0$ and $N\in(\tilde{Z_-},
\tilde{Z_+})$, while $Z<0$ for $f>0$ and $N\in(0,\tilde{Z_-})\bigcup(\tilde{Z_+},\infty)$,
as well as for $f<0$ and any $N$.

In addition, we should provide the condition $\varphi^2>N\psi$ to
fulfil. Two possibilities exist to satisfy this condition,
depending on the signs of $\mu_j$:
\begin{enumerate}
\item for $\mu_1>0$ and $\mu_2>0$ and any $\alpha$ we should assume the
restriction $N<B$ to fulfil, where
\be \label{eq18}
B=\frac{1}{4}\frac{d_1^2+d_2^2}{\mu_1u_0^2+\mu_2v_0^2};
\ee

\item there are no restrictions on $N$ when $\mu_1<0$ and $\mu_2<0$ for
generic $u_0$ and $v_0$ and any $\alpha$.
\end{enumerate}
On the other hand, $\varphi^2<N\psi$ for $\mu_1>0$, $\mu_2>0$ and
$N>B$.

\subsection{Summary of the analysis}
Summarizing the results obtained from the analysis of solutions $\Omega_\pm^2$
(\ref{eq12}), we infer that stability regions ($D>0$, $\sqrt{D}<\varphi^2-N\Psi$)
are determined by the following conditions on the parameters of equations (\ref{CEq})
for generic choice of the amplitudes $u_0$ and $v_0$:
\begin{eqnarray}\label{eq19}\eqalign{
\mathrm{(a)}\   \mu_1<0,\,\mu_2<0,\,\alpha<1,\,\forall N;\\
\mathrm{(b)}\   \mu_1>0,\,\mu_2>0,\,\alpha>1,\,N\in(0,\min(B,Z_+));\\
\mathrm{(c)}\   \mu_1<0,\,\mu_2<0,\,\alpha>1,\,N\in(0,Z'_+);\\
\mathrm{(d)}\  \mu_1>0,\,\mu_2>0,\,\alpha<1,\,\left\{\begin{array}{cc}
\displaystyle{N\in(0, \min(B,\tilde{Z_-}))}, \;&
B<\tilde{Z_+},\vspace{1mm}\\ N\in(0,\tilde{Z_-})\bigcup(\tilde{Z_+}, B),
\;& \mathrm{if} \;
 \tilde{Z_+}<B.
 \end{array}\right.
}\end{eqnarray}

Accordingly, instability regions are characterized by the parameters
\begin{eqnarray}\label{eq20}\eqalign{
\mathrm{(e)}\   \mu_1<0,\,\mu_2>0,\,\alpha<1,\,N\in(N'_-,N'_+);\\
\mathrm{(f)}\   \mu_1>0,\,\mu_2>0,\,\alpha>1,\,N>\min(B,Z_+);\\
\mathrm{(g)}\   \mu_1<0,\,\mu_2<0,\,\alpha>1,\,N>Z'_+;\\
\mathrm{(h)}\
\mu_1>0,\,\mu_2>0,\,\alpha<1,\,\left\{\begin{array}{cc}
\displaystyle {N\in(\tilde{Z_-}, \tilde{Z_+})}, \;&\mathrm{if}\;
B<\tilde{Z_-},\qquad\;\vspace{1mm}\\
N\in(B,\tilde{Z_+}),\; \;& \mathrm{if} \; B\in(\tilde{Z_-},\tilde{Z_+}),\\
N>B,\qquad\;\; \;& \mathrm{if}\; B>\tilde{Z_+}.\qquad\;
 \end{array}\right.
}\end{eqnarray}

It should be noted that stability properties depend on the sign of
the products $\mu_j=s_jd_j$ and not on $s_j$ and $d_j$ separately.
Note also that we reproduce the results of Kourakis and
Shukla~\cite{rev16} in the local limit ($N=\kappa^{-2}$). At the
same time, the availability of an additional degree of freedom in
the form of nonlocality enables us to manage the conditions of
stability/instability. In the next subsection we demonstrate this
possibility.

\subsection{Examples}
\begin{figure}[]
\begin{center}
 \includegraphics[width=0.5\textwidth]{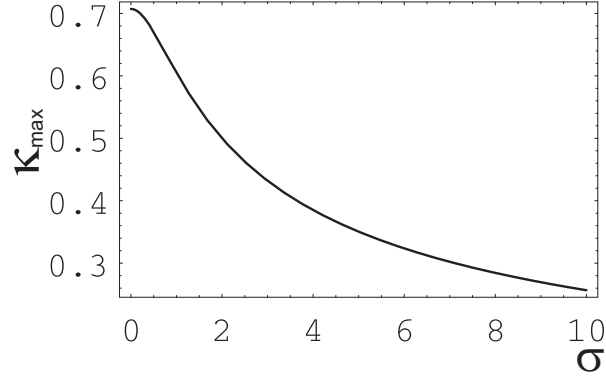}
\caption{Maximum wave number $\kappa_{\mathrm{max}}$ of the
instability range $(0,\,\kappa_{\mathrm{max}})$ versus the
nonlocality parameter $\sigma$. Nonlocality suppresses the
instability bandwidth. The instability regime corresponds to the
case (g) in (\ref{eq20}).} \label{fig1}
\end{center}
\end{figure}
The instability growth rate is characterized by the quantity \be
\label{eq21}
\lambda(\kappa)=\max\left(\left|\mathrm{Im}\left(\sqrt{\Omega_-^2}\right)\right|,
\left|\mathrm{Im}\left(\sqrt{\Omega_+^2}\right)\right|\right). \ee

\begin{figure}
\begin{center}
 \includegraphics[width=0.5\textwidth]{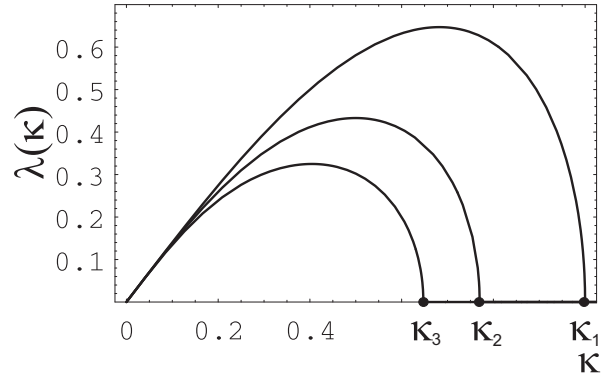}
\caption{MI gain spectra for the instability regime (g). The
points $\kappa_1$, $\kappa_2$ and $\kappa_3$ corresponds to
$\kappa_{\mathrm{max}}$ in figure \ref{fig1} for $\sigma=1$, $2$,
and $3$, respectively.} \label{fig2}
\end{center}
\end{figure}

\begin{figure}
\begin{center}
 \includegraphics[width=0.5\textwidth]{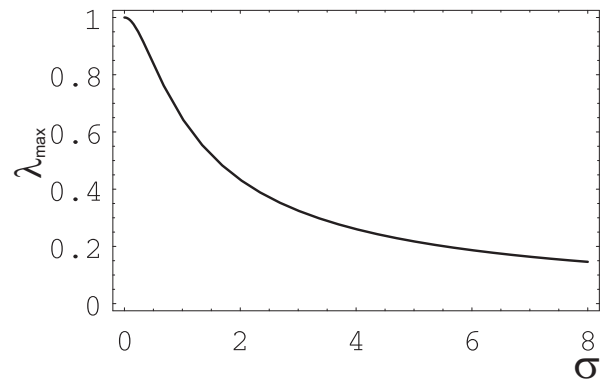}
\caption{Dependence of the MI gain maximum
$\lambda_{\mathrm{max}}$ on the nonlocality parameter $\sigma$ for
the instability regime (g).} \label{fig3}
\end{center}
\end{figure}

Below we use the exponential response function (\ref{exp}). For
this function we can invert the dependence $N(\kappa)$:
\be\label{eq22}
\kappa^2=\frac{1}{2\sigma^2}\left(\sqrt{1+\frac{4\sigma^2}{N}}-1\right)
\ee We fix magnitudes of all the parameters, except for $\sigma$,
to reveal the role of nonlocality in the development of MI for a
number of regimes listed in (\ref{eq20}). As regards the
cross-phase modulation parameter $\alpha$, we take its featured
values, namely, $\alpha=2$ for $\alpha>1$ and $\alpha=2/3$ for
$\alpha<1$. At first we consider the regime (g) in (\ref{eq20})
and choose the parameters as $\mu_1=\mu_2=-1$, $u_0=v_0=1$,
$d_1^2=d_2^2=4$, $\alpha=2$. They correspond to the defocusing
nonlinearity in planar waveguides or normal dispersion in a
focusing Kerr medium (recall that MI does not occur in such
conditions for a single field). Since $N>Z'_+$ for the regime (g),
we have $\kappa\in(0,\kappa_{\mathrm{max}})$ from (\ref{eq17})
with
\[
 \kappa_{\mathrm{max}}^2=\frac{1}{2\sigma^2}\left(\sqrt{1+\frac{4\sigma^2}{Z'_+}}-1\right).
\]
Figure \ref{fig1} shows the dependence
$\kappa_{\mathrm{max}}(\sigma)$ with $Z'_+$ from (\ref{eq17}). We
see that the range $(0,\kappa_{\mathrm{max}})$ of instability wave
numbers narrows with growing nonlocality. Figure \ref{fig2} shows
the MI gain spectra $\lambda(\kappa)$ (\ref{eq21}) for three
values of the nonlocality parameter $\sigma$. The maximum gain
$\lambda_{\mathrm{max}}$ decreases with an increase in the
nonlocality parameter $\sigma$. Such a decrease is illustrated by
figure \ref{fig3}.

\begin{figure}
\begin{center}
 \includegraphics[width=0.5\textwidth]{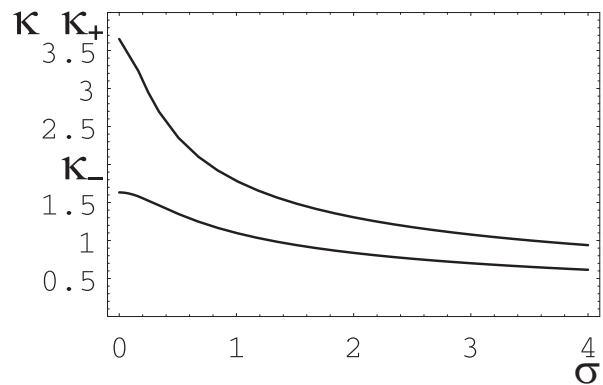}
\caption{Dependence of limiting values $\kappa_\pm$ of modulation
wave numbers on the nonlocality parameter $\sigma$ for the
instability regime (e). The instability bandwidth shrinks with
increasing nonlocality.} \label{fig4}
\end{center}
\end{figure}

\begin{figure}
\begin{center}
 \includegraphics[width=0.5\textwidth]{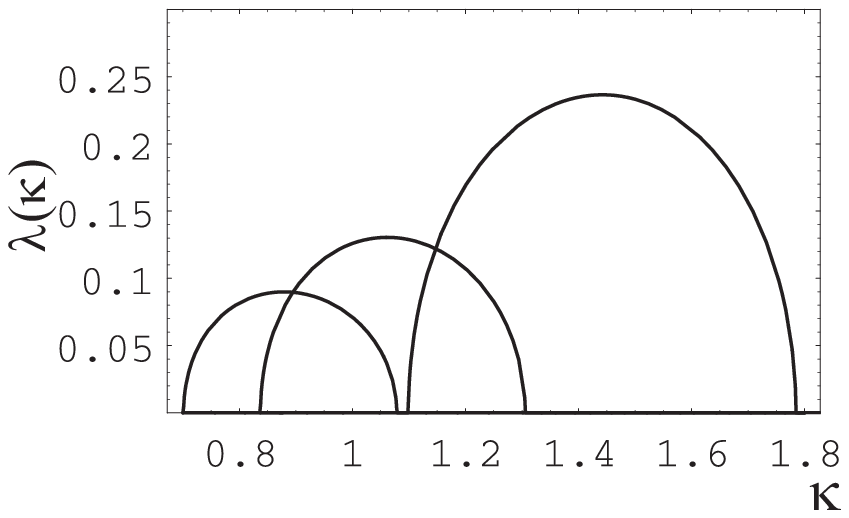}
\caption{MI gain spectra for the instability regime (e) for
different values of nonlocality parameter $\sigma$. Three curves
correspond (from right to left) to $\sigma=1$, 2, 3.} \label{fig5}
\end{center}
\end{figure}

The second example covers the regime (e) in (\ref{eq20}). This
regime describes a joint propagation of two waves, when one of
them `sees' the normal dispersion of a medium, while the second
wave moves in the anomalous dispersion environment. Since the
allowable values of $N$ lie within the interval ($N_-',N_+'$), we
have two limiting values of the modulation wave numbers, namely,
$\kappa_-$ and $\kappa_+$:
\[
\kappa_\pm=\frac{1}{2\sigma^2}\left(\sqrt{1+\frac{4\sigma^2}{N'_\mp}}-1\right)
\]
We choose $\mu_1=-1$, $\mu_2=1$, $u_0=v_0=1$, $d_1^2=2$,
$d_2^2=4$, $\alpha=2/3$. As it is seen from figure \ref{fig4}, the
interval width $(\kappa_-,\kappa_+)$ decreases with increasing
$\sigma$. Owing to the dependence of $\kappa_\pm$ on the
nonlocality parameter $\sigma$, the MI gain spectra demonstrate
different boundary values of the bandwidth position, both minimal
and maximal, for different $\sigma$ (figure \ref{fig5}).

\section{Group-velocity mismatch and nonlocality management}

When we account for group-velocity mismatch $(V_1\neq V_2)$, the
dispersion relation (\ref{eq6}) gives a general fourth-order
algebraic equation. To avoid cumbersome calculation, we will
follow \cite{rev27,rev16} and give a geometrical description of
the nonlocality management in the case of the group-velocity
mismatch.

Let us write the dispersion relation (\ref{eq6}) in the form \be
\label{eq23} \left(\Omega-V_1\kappa
\right)^2-\Omega_1^2=\frac{\Omega_{12}^4}{ \left(\Omega-V_2\kappa
\right)^2-\Omega_2^2}. \ee This formula describes a mutual
arrangement of two curves -- a parabola on the left and a more
complicated `right' curve. We know from the previous exposition
that the stability regime is provided by four real solutions of
the dispersion equation (\ref{eq6}), or, in other words, by four
intersection points of the above curves. If the number of
intersection points is less than four, the system is unstable. We
show below that nonlocality of a medium provides an efficient tool
to manage stability/instability properties giving a possibility to
assure stability of a system which is unstable in a local medium.

\begin{figure}
\begin{center}
 \includegraphics[width=0.5\textwidth]{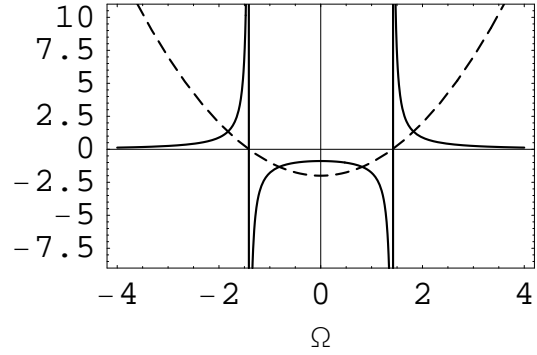}
\caption{Zero group-velocity mismatch ($V_1=V_2$): four
intersection points of the curves (the parabola is depicted by
dashed line) defined in (13) provide stability of the continuous
wave solution for a local medium ($\sigma=0$) with the parameters
$d_1^2=d_2^2=4$, $\mu_1=\mu_2=1$, $\kappa^2=2$,
$u_0^2=v_0^2=1$.}\label{fig6}
\end{center}
\end{figure}

\begin{figure}
\begin{center}
 \includegraphics[width=0.5\textwidth]{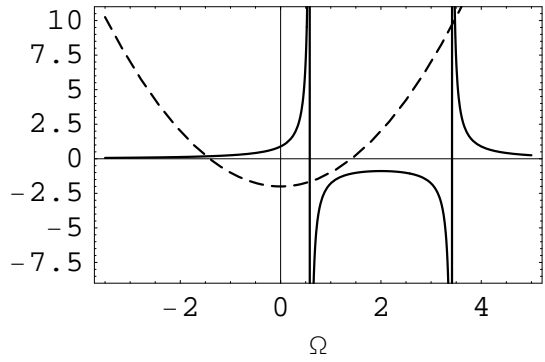}
\caption{Breaking of stability (two intersection points only) in a
local medium ($\sigma=0$) caused by nonzero group-velocity
mismatch ($V_1=0$, $V_2=\sqrt{2}$). Other parameters are the same
as in figure \ref{fig6}.}\label{fig7}
\end{center}
\end{figure}

\begin{figure}
\begin{center}
 \includegraphics[width=0.5\textwidth]{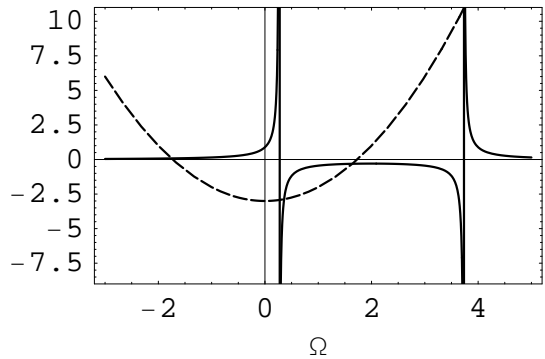}
\caption{Restoration of stability (four intersection points) in a
nonlocal medium ($\sigma=1$) with nonzero group-velocity mismatch
($V_1=0$, $V_2=\sqrt{2}$). Other parameters are the same as in
figure \ref{fig6}.}\label{fig8}
\end{center}
\end{figure}

Figure \ref{fig6} corresponds to light wave propagation in a local
medium $(\sigma=0)$ with the parameters $\mu_1=\mu_2=1$,
$u_0=v_0=1$, $d_1^2=d_2^2=4$, $\alpha=2/3$, $\kappa^2=2$, and
$V_1=V_2=0$, i.e., zero group-velocity mismatch. The choice of
parameters gives $\Omega_1^2=\Omega_2^2=2$ and
$\Omega_{12}^4=16/9$ and means individual stability of both waves.
There are four real intersection points of the parabola with the
`right' curve, the latter consists of two hyperbolae and a curve
in the lower half plane with two singular points. Therefore, this
situation corresponds to absolute stability. Let us now introduce
the velocity mismatch, taking $V_1=0$ and $V_2=\sqrt{2}$. The
`right' curve shifts horizontally, losing two intersection points
and thereby producing instability (figure \ref{fig7}). If,
however, nonlocality is coming into play [we take $\sigma=1$ and
recalculate $\Omega_1^2$, $\Omega_2^2$ and $\Omega_{12}^4$ in
accordance with (\ref{eq7})], the four intersection points occur
again, and stability is restored (figure \ref{fig8}).

\begin{figure}
\begin{center}
 \includegraphics[width=0.45\textwidth]{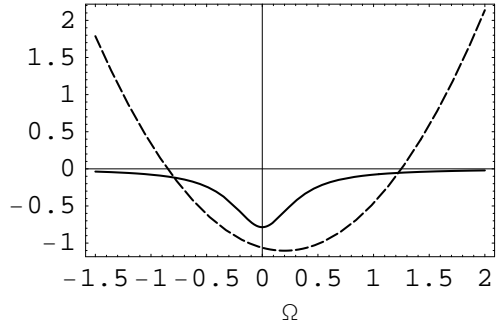}
\caption{Stable-unstable wave pair in a local medium with the
group-velocity mismatch. The parameters are $\mu_1=-1.5$,
$\mu_2=1$, $u_0^2=v_0^2=0.6$, $\alpha=2/3$, $\kappa^2=1$,
$V_1=0.2$, $V_2=0$. The system is unstable because there are only
two intersection points of both curves.}\label{fig9}
\end{center}
\end{figure}

\begin{figure}[]
\begin{center}
 \includegraphics[width=0.5\textwidth]{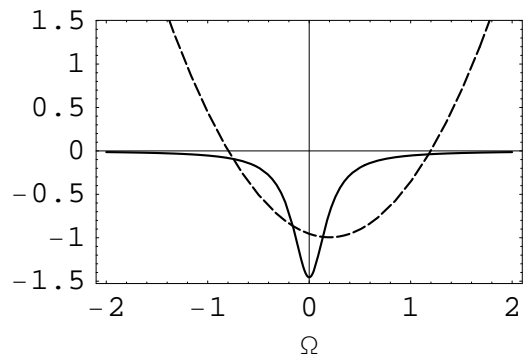}
\caption{Restoration of stability of the stable-unstable wave pair
in a nonlocal medium ($\sigma=0.5$). Other parameters are the same
as in figure \ref{fig9}.}\label{fig10}
\end{center}
\end{figure}

Our second example is concerned with a stable-unstable wave pair
determined by the parameters $\mu_1=-1.5$, $\mu_2=1$,
$u_0=v_0=0.6$, $\alpha=2/3$, $\kappa^2=1$ which correspond to
$\Omega_1^2=1.1025$, $\Omega_2^2=-0.11$ and
$\Omega_{12}^4=-0.0864$. For the group-velocity mismatch
($V_1=0.2$, $V_2=0$) there are only two intersection points -- the
system is unstable (figure \ref{fig9}). Retaining this mismatch
but allowing nonzero nonlocality ($\sigma=0.5$), we restore the
system stability (figure \ref{fig10}).

\begin{figure}[]
\begin{center}
 \includegraphics[width=0.5\textwidth]{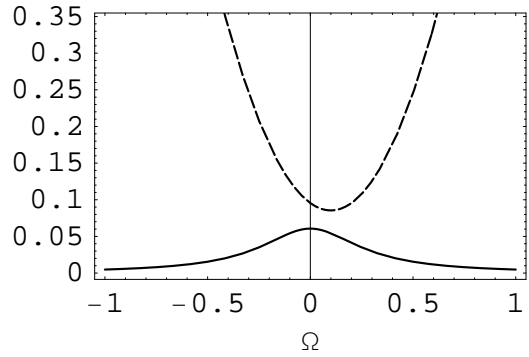}
\caption{Unstable-unstable wave pair in a local medium. The
parameters are $\mu_1=\mu_2=0.3$, $u_0=v_0=0.6$, $\alpha=2/3$,
$\kappa^2=1$, $V_1=0.1$, $V_2=0$ which correspond to
$\Omega_1^2=\Omega_2^2=-0.0855$. There are no intersection points,
the system is absolutely unstable.}\label{fig11}
\end{center}
\end{figure}

\begin{figure}[]
\begin{center}
 \includegraphics[width=0.5\textwidth]{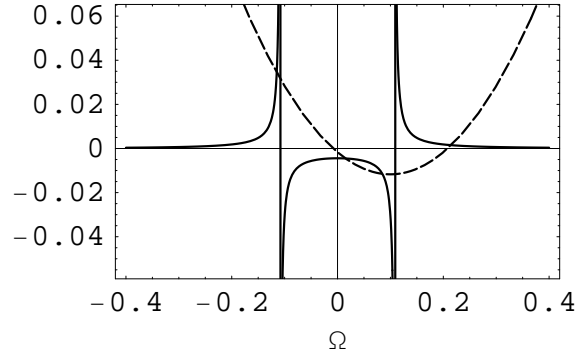}
\caption{Unstable-unstable wave pair in a nonlocal medium
($\sigma=3$, $\Omega_1^2=\Omega_2^2=0.0117$). Other parameters are
the same as in figure \ref{fig11}. The system is stable at the
cost of nonlocality management.}\label{fig12}
\end{center}
\end{figure}

 The last example provides the most emphatic illustration of the nonlocality management.
 Let us consider the unstable-unstable wave pair which corresponds to negative values of
 $\Omega_1^2$ and $\Omega_2^2$. It is stated in \cite{rev16} that such a pair
 is \textit{always unstable} in a local medium (see figure \ref{fig11}, no intersection points).
 The situation is drastically altered in a nonlocal medium. It is seen from (\ref{eq7})
 that $\Omega_j^2$ can change sign for sufficiently large nonlocality parameter
 $\sigma$, providing stability of the system. This possibility is exemplified by
 figure \ref{fig12}.


 \section{Conclusion}
 We have investigated the MI of two interacting waves in a nonlocal Kerr medium, both
 for zero and nonzero group-velocity mismatch. The case of zero mismatch admits the
 full analytical description of stable/unstable regimes of the wave propagation. We have
 demonstrated that nonlocality suppresses the growth rate peak and bandwidth of instability.
 For the case of nonzero velocity mismatch we have revealed a crucial role of nonlocality
 in management of the stability properties in the system under consideration. In particular,
 we can provide stability of the nonlocal system in a regime when its local counterpart
 demonstrates absolute instability.

\end{document}